\documentclass[12pt]{article}
\pdfoutput=1

\usepackage{epsfig,latexsym,amsfonts,amsmath,amsthm,amssymb,amsbsy,multirow,slashed,wasysym,textcomp,comment,mathtools,cancel,cite}

 \newcommand{\badat}{\begin{alignedat}}
 \newcommand{\eadat}{\end{alignedat}}

\long\def\new#1\endnew{{\bf #1}}		
\long\def\del#1\enddel{}

\def\del{\partial}
\def\nn{\nonumber}

\usepackage{color}

\newcommand{\pink}[1]{\textcolor{\pink}{#1}}

\definecolor{dblue}{rgb}{0.2,0.50,0.80}

\def\A{\mathcal{A}}

\def\H{\mathcal{H}}

\def\bh{{\bar h}}
\def\bz{{\bar z}}

\def\ba{{\bar a}}
\def\bb{{\bar b}}
\def\bc{{\bar c}}
\def\bd{{\bar d}}

\numberwithin{equation}{section} 

\begin{document}

 \begin{titlepage}
  \thispagestyle{empty}
  \begin{flushright}
  CPHT-RR021.052019 
  \end{flushright}
  \bigskip
  \begin{center}
	 \vskip2cm
  \baselineskip=13pt {\LARGE \scshape{Conformally Soft Theorem In Gravity}}
	 \vskip1.5cm
   \centerline{
   {Andrea Puhm}
   {}
   }
	
 \bigskip\bigskip\bigskip

\centerline{\em CPHT, CNRS, Ecole polytechnique, IP Paris,}
\centerline{\em F-91128 Palaiseau, France}

\bigskip

\centerline{\em Black Hole Initiative, Harvard University,}
\centerline{\em Cambridge, MA 02138, USA}

\bigskip

\centerline{\em Center for the Fundamental Laws of Nature, Harvard University,}
\centerline{\em Cambridge, MA 02138, USA}

 \bigskip\bigskip
 
  \end{center}

\begin{abstract}
  \noindent
	
A central feature of scattering amplitudes in gravity or gauge theory is the existence of a variety of energetically soft theorems which put constraints on the amplitudes. Celestial amplitudes which are obtained from momentum-space amplitudes by a Mellin transform over the external particle energies cannot obey the usual energetically soft theorems. Instead, the symmetries of the celestial sphere imply that the scattering of {\it conformally soft} particles whose conformal weights under the 4D Lorentz group $SL(2,\mathbb{C})$ are taken to zero obey special relations. Such conformally soft theorems have recently been found for gauge theory. Here, I show conformally soft factorization of celestial amplitudes for gravity and identify it as the celestial analogue of Weinberg's soft graviton theorem. 

\end{abstract}

 \end{titlepage}
\tableofcontents
\section{Introduction}

The 4D Lorentz group $SL(2,\mathbb{C})$ acts as the global conformal group on the celestial two-sphere at null infinity where massless asymptotic scattering states are defined. Scattering amplitudes are usually discussed in a momentum basis where translation invariance is manifest but conformal properties are hidden. ``Celestial amplitudes'' which are obtained from the usual momentum-space amplitudes by a Mellin transform over the external particle energies obscure translation symmetry but render the conformal action trivial. In this conformal basis asymptotic states are labelled by their $SL(2,\mathbb{C})$ Lorentz/conformal weights $(h, \bh)$ (or equivalently their total conformal dimension $\Delta=h+\bh$ and spin $J=h-\bh$) rather than the usual energy-momentum four-vector.

In~\cite{Pasterski:2017kqt} a basis for flat space amplitudes was constructed in terms of conformal primary wavefunctions with total conformal dimensions in the unitary principal series of the Lorentz group $\Delta=1+i\lambda$ with $\lambda\in \mathbb{R}$. An important subtlety arises for the zero-modes $(\lambda=0$) which were not considered in~\cite{Pasterski:2017kqt} but were explicitly constructed in~\cite{Donnay:2018neh}; when these modes are included the conformal primary wavefunctions on the unitary principal series form a complete $\delta$-function normalizable basis for flat space amplitudes.

A central feature of the usual scattering amplitudes in gravity or gauge theory is the existence of a variety of (energetically) soft factorization theorems which put constraints on the amplitudes. However, in the conformal basis the notion of a soft particle is lost. $SL(2,\mathbb{C})$ primary wavefunctions are not energy eigenstates so the energy cannot be taken to zero. Instead, we have the notion of a {\it conformally soft} particle for which the conformal weights either $h$ or $\bh$ is taken to zero~\cite{Cheung:2016iub,Donnay:2018neh}. The symmetries of the celestial sphere imply that the scattering of such particles also obey special relations.

Recently, there has been considerable interest in studying scattering amplitudes in the conformal basis. 
Celestial amplitudes for Yang-Mills theory at tree-level were recently constructed in~\cite{Pasterski:2017ylz,Schreiber:2017jsr}.
In theories with sufficiently soft UV behavior, tree-level celestial gluon scattering obeys {\it conformally soft} theorems~\cite{Fan:2019emx,Nandan:2019jas,Pate:2019mfs} involving $h\to 0$ or $\bh \to 0$. This is both expected and suprising. MHV amplitudes obey soft theorems which are equivalent to the Ward identities of spontaneously broken large gauge symmetries~\cite{He:2015zea}. The generators of these symmetries correspond to 2D Kac-Moody currents with $(h,\bh)=(1,0)$ or $(0,1)$ on the celestial sphere~\cite{He:2015zea,Donnay:2018neh} which can be understood as the {\it conformally soft} $\lambda\to 0$ limit of spin~1 primaries with conformal weights $(h,\bh)=(1+\frac{i\lambda}{2},\frac{i\lambda}{2})$ or $(\frac{i\lambda}{2},1+\frac{i\lambda}{2})$~\cite{Donnay:2018neh}. Hence the insertion of conformally soft currents into celestial correlators is expected to give rise to the celestial analogue of the soft theorems. On the other hand, it is suprising because celestial amplitudes involve a superposition of all energies and so, unlike the energetically soft theorems, the conformally soft theorems cannot be derived from low-energy effective field theory~\cite{Pate:2019mfs}. Nevertheless, it was shown~\cite{Fan:2019emx,Nandan:2019jas,Pate:2019mfs} that the $\lambda\to0$ limit of celestial gluon amplitudes reproduces the well-known energetically soft factor of gluon scattering.

The situation is even more puzzling in gravity. 
Weinberg's soft graviton theorem~\cite{Weinberg:1965nx} can be understood as the Ward identity of spontaneously broken BMS supertranslation symmetry~\cite{He:2014laa}. The generator of this symmetry, the 2D supertranslation current\cite{Strominger:2013jfa}, can be understood from the divergence of the $\lambda \to 0$ limit of a spin 2 primary operator on the celestial sphere with conformal weights $(h,\bh)=(\frac{3}{2}+\frac{i\lambda}{2},-\frac{1}{2}+\frac{i\lambda}{2})$ or $(-\frac{1}{2}+\frac{i\lambda}{2},\frac{3}{2}+\frac{i\lambda}{2})$~\cite{Donnay:2018neh}. Unlike the soft photon or gluon current in gauge theory, the supertranslation current is {\it not conformally soft} as defined above, but instead has conformal weights $(h,\bh)=(\frac{3}{2},\frac{1}{2})$ or $(\frac{1}{2},\frac{3}{2})$. Moreover, its OPE with another operator shifts the conformal weights of the latter by $(\frac{1}{2},\frac{1}{2})$~\cite{Donnay:2018neh}. Supertranslation invariance thus appears to provide an infinite number of new constraints in the conformal basis which recursively relate operators with different conformal dimension. 

In the language of amplitudes, one would expect to find a celestial analogue of Weinberg's soft theorem in the conformal basis. However, in Einstein gravity, the Mellin transforms diverge and hence the classical four-graviton celestial amplitudes do not exist~\cite{Stieberger:2018edy}. This uncontrollable growth becomes supersoft, exponentially suppressed at high energies, in string theory. 
Having to resort to string theory for the purpose of studying the conformally soft behavior of amplitudes appears superfluous. On the other hand, the Mellin transforms do involve a superposition of all energies and thus mix the UV and the IR in the conformal basis.

The purpose of this paper is two-fold. First, I will argue that even classical celestial graviton amplitudes exist for a certain analytic continuation in $\lambda$ of the conformal dimensions of some external gravitons away from the principal continuous series\footnote{I would like to thank Agnese Bissi for collaboration on a related project~\cite{BP19} from which this argument arose.}. The reason for analytically continuing $\lambda$ is not ad hoc but is implied by conformal covariance of the amplitude. Moreover, in~\cite{Donnay:2018neh} we showed that a conformally soft spin~2 mode with dimension $\Delta=2$ is related to the 2D stress tensor for 4D gravity~\cite{Kapec:2016jld}. This mode is obtained from a general spin~2 conformal primary in the construction of~\cite{Pasterski:2017kqt} by setting $\lambda=-i$. Hence, primaries away from the principal continuous series may have to be included for a complete holographic description of 4D quantum gravity\footnote{It is conceivable that these primaries may be obtainable by a suitable contour deformation of a convolution of primaries on the principal continuous series; a thorough understanding of this point is left for future work. I would like to thank Sabrina Pasterski and Andy Strominger for discussion.}.

Second, I will show that the celestial analogue of Weinberg's soft theorem can be understood as the $\lambda\to 0$ limit of celestial graviton amplitudes with the feature that higher-point amplitudes are related to lower-point amplitudes with shifted conformal dimensions.
This limit is not conformally soft as defined above. Nevertheless, I will refer to the celestial version of Weinberg's soft theorem as the {\it conformally soft graviton theorem}. This is only a slight abuse of language as the combined action of the insertion of the supertranslation current with weights $(\frac{3}{2},\frac{1}{2})$ or $(\frac{1}{2},\frac{3}{2})$ and the shift in weights by $(\frac{1}{2},\frac{1}{2})$ that it induces may be thought of as conformally soft.

This paper is organized as follows. In section~\ref{celestialH}, I introduce celestial amplitudes for gravity and argue that analytic continuation of $\lambda$ to complex values allows to interpret the otherwise divergent energy integral as a distribution. I review celestial three- and four-graviton amplitudes constructed first in~\cite{Stieberger:2018edy} in sections~\ref{celestialH3point} and~\ref{celestialH4point} respectively.
In section~\ref{csoftgravity}, I present a general argument that translates Weinberg's soft factor into the conformal basis. The conformally soft theorem is explicitly verified in section~\ref{csoftgravity4point} for the known tree-level celestial four-graviton amplitudes, and for four-graviton heterotic string amplitudes in section~\ref{csoftgravity4pointHeterotic}. In section~\ref{csoftgravitynpoint}, I present an argument that extends the conformally soft theorem to $n$-graviton MHV amplitudes.

{\bf Note added:} While this paper was being prepared for submission the preprint~\cite{AMS} appeared which studies celestial amplitudes using ambitwistor strings and has overlapping results. After accounting for different conventions their formula for the conformally soft limit of celestial gravity amplitudes appears to agree with~\eqref{csoftgravitynpoint} below.

\section{Celestial amplitudes}\label{celestialH}

\subsection{Celestial $n$-gluon and $n$-graviton amplitudes}
Consider an $n$-point scattering amplitude
\begin{equation}\label{An}
 \A_{\ell_1\dots\ell_n}(\omega_i;z_i,\bz_i)=A_{\ell_1\dots\ell_n}(\omega_i;z_i,\bz_i)\delta^{(4)}\Big(\sum_{i=1}^n p^\mu_i \Big)\,,
\end{equation}
where $\ell_i$ labels the helicity, and I parametrize a null four-momentum by a sign $\epsilon_i=\pm1$ (for outgoing and ingoing particles respectively), a positive frequency $\omega_i$ and a point $(z_i,\bz_i)$ on the celestial sphere such that $p^\mu_i=\epsilon_i \omega_i q^\mu_i(z_i,\bz_i)$ with
\begin{equation} \label{qmppp}
 q^\mu_i(z_i,\bz_i)=(1+z_i \bz_i, z_i+\bz_i, -i(z_i-\bz_i), 1-z_i\bz_i)\,.
\end{equation}
The ``celestial amplitude'' is obtained from the standard momentum-space amplitude~\eqref{An} by a Mellin transform on each of the external particles 
\begin{equation} \label{celesitalA}
 \widetilde{\A}_{J_1\dots J_n}(\lambda_i;z_i,\bz_i)=\prod_{k=1}^n \Big(\int_0^\infty d\omega_k \omega_k^{i \lambda_i}\Big) \A_{\ell_1\dots\ell_n}(\omega_i;z_i,\bz_i)\,.
\end{equation}
One may show~\cite{Pasterski:2017ylz,Pasterski:2017kqt} that under $SL(2,\mathbb{C})$ Lorentz transformations
\begin{equation}
  \widetilde{\A}_{J_1\dots J_n}\left(\lambda_i,\frac{a z_i+b}{cz_i+d},\frac{ \ba  \bz_i+ \bb}{ \bc \bz_i+ \bd}\right) =\prod_{j=1}^n \Big[ (c z_j+d)^{\Delta_j+J_j} (\bc \bz_j+\bd)^{\Delta_j-J_j}\Big] \widetilde{\A}_{J_1\dots J_n}(\lambda_i;z_i,\bz_i)\,,
\end{equation}
with conformal dimensions $\Delta_i=1+i\lambda_i$ and spins~$J_i\equiv \ell_i$, which in turn are related to $(h_i,\bh_i)$ by $(h_i,\bh_i)=\frac{1}{2}(\Delta_i+J_i,\Delta_i-J_i)$. Celestial amplitudes therefore share conformal properties with correlation functions on the celestial sphere.
The momentum space amplitudes are conventionally normalized with inner product
\begin{equation}
 (p_1 \ell_1; p_2 \ell_2)=(2\pi)^3 2p_1^0 \delta^{(3)}(\vec{p}_1+\vec{p}_2)\delta_{\ell_1,-\ell_2}\,.
\end{equation}
which implies the celestial inner product
\begin{equation}
 (\lambda_1, z_1 ,\bz_1,J_1;\lambda_2, z_2, \bz_2,J_2)=(2\pi)^4 \delta(\lambda_1+\lambda_2) \delta^{(2)}(z_1-z_2) \delta_{J_1,-J_2}\,.
\end{equation}

Celestial amplitudes for Yang-Mills theory have been constructed at tree-level for three and four gluons in~\cite{Pasterski:2017ylz} and were generalized to $n$ gluons in~\cite{Schreiber:2017jsr}. In theories with sufficiently soft UV
behavior, tree-level celestial gluon scattering obeys {\it conformally soft} theorems~\cite{Fan:2019emx,Nandan:2019jas,Pate:2019mfs} involving $h\to 0$ or $\bh \to 0$. 
Here I will be interested in tree-level celestial graviton scattering amplitudes, which I will denote by $\H$ to distinguish them from gluon amplitudes $\A$, and their conformally soft behavior. In the following, powers of the gauge/gravitational (or string) coupling are absorbed into the wave function normalization.

In momentum-space the general $n$-point graviton amplitude is given by
\begin{equation}\label{Hn}
 \H_{\ell_1\dots\ell_n}(\omega_i;z_i,\bz_i)=H_{\ell_1\dots\ell_n}(\omega_i;z_i,\bz_i)\, \delta^{(4)}\Big(\sum_{i=1}^n \epsilon_i \omega_i q_i(z_i,\bz_i) \Big)\,.
\end{equation}
The spinor-helicity formalism is a convenient and powerful framework for expressing amplitudes. I use the notation
\begin{equation}
 \langle ij \rangle =-2\epsilon_i \epsilon_j \sqrt{\omega_i \omega_j} z_{ij}\,,\quad
 [ij] =2 \sqrt{\omega_i \omega_j} \bz_{ij}\,,\\
\end{equation}
where $z_{ij}\equiv z_i-z_j$ and $\bz_{ij}=\bz_i-\bz_j$ and scalar products are defined as
\begin{equation}
 s_{ij}\equiv -2\, p_i \cdot p_j = -\langle ij\rangle [ij]=4 \epsilon_i \epsilon_j \omega_i \omega_j z_{ij} \bz_{ij}\,.
\end{equation}
A by now celebrated result is that the stripped amplitude $H_{\ell_1\dots\ell_n}$ can be conveniently expressed by sums of squares of stripped gluon amplitudes $A_{\ell_1\dots\ell_n}$~\cite{Kawai:1985xq,Bern:2008qj}weighted by the kinematic invariants. The general formula for the stripped $n$-point MHV Yang-Mills amplitude in momentum space ($\ell_1=\ell_2=-1$ and $\ell_3=\dots=\ell_n=+1$) is\cite{Parke:1986gb}
\begin{equation}\label{Angluon}
\badat{2}
 A_{--+\dots +}(\omega_i;z_i,\bz_i)&=\frac{\langle12\rangle^3}{\langle 23 \rangle \cdots \langle n 1 \rangle}\\
 &\equiv A_n(1^-2^-3^+\dots n^+)\,.
\eadat
\end{equation} 
The stripped MHV $n$-graviton amplitude in momentum space can be expressed as~\cite{Elvang:2007sg}
\begin{equation}\label{HnElvangFreedman}
\badat{2}
 H_{--+\dots +}(\omega_i;z_i,\bz_i)&\stackrel{n\geq 4}{=} \sum_{\mathcal{P}(i_3\dots i_n)} s_{1\, i_n} \left(\prod_{m=4}^{n-1} \beta_m\right) \Big[A_n(1^-2^-i_{3}^+\dots i_{n}^+)\Big]^2\\
 &\equiv  H_n(1^-2^-3^+\dots n^+)\,,
\eadat
\end{equation}
where the sum is over all permutations $\mathcal{P}(i_3\dots i_n)$ of the external positive helicity labels $\{3,\dots,n\}$ and 
\begin{equation}
 \beta_m= -\frac{\langle i_m\, i_{m+1}\rangle}{\langle 2 \, i_{m+1}\rangle} \langle 2|i_3 +\dots +i_{m-1}|i_m]\,,
\end{equation}
where 
\begin{equation}
\langle i|k|j\rangle=\langle i k\rangle [kj]\,. 
\end{equation}

The celestial $n$-graviton amplitude is obtained by the following Mellin transform
\begin{equation}\label{celestialHn}
\badat{3}
 \widetilde{\H}_{--+\dots+}&=\left(\prod_{k=1}^n \int_0^\infty d\omega_k \omega_i^{i \lambda_k} \right) \H_{--+\dots +}\\
 &=\frac{1}{(-2)^{2n-8}}\left(\prod_{k=1}^n \int_0^\infty d\omega_k \omega_k^{i \lambda_k-J_k} \right) \left[ \sum_{\mathcal{P}(i_3 \dots i_n)} s_{1  i_n} \left( \prod_{m=4}^{n-1} \beta_m\right) \left( \frac{z_{12}^3}{z_{2 i_3} \dots z_{i_n 1}}\right)^2\right]\\
 &\quad \times \delta^{(4)}\left(\sum_{i=1}^n \epsilon_i \omega_i q_i\right)\,.
\eadat
\end{equation}
In section~\ref{csoftgravity4point}, I will present the details of Weinberg's soft theorem in the conformal basis for celestial three- and four-graviton amplitudes, first constructed in~\cite{Stieberger:2018edy}, which I will now review.

\subsection{Celestial three-graviton amplitude}\label{celestialH3point}

To study the three-graviton amplitude one resorts to $(2, 2)$ signature where the amplitude
is non-vanishing. Lorentz transformations act as $SL(2,\mathbb{R})\times SL(2,\mathbb{R})$  on $z_i$ and $\bz_i$ which now are independent real variables.
The MHV graviton three-point amplitude ($\ell_1=\ell_2=-1$ and $\ell_3=+1$) is given by 
\begin{equation}\label{H3}
 \H_{--+}(p_i)=H_{--+}\, \delta^{(4)}\Big(\sum_{i=1}^3 \epsilon_i \omega_i q_i\Big)\,,
\end{equation}
where
\begin{equation}
 H_{--+}=A_{--+}^2=\left(\frac{\omega_1 \omega_2}{\omega_3} \frac{z_{12}^3}{z_{23}z_{31}}\right)^2\,.
\end{equation}
and a convenient way of writing the momentum-conserving $\delta$-function is~\cite{Pate:2019mfs}
 \begin{equation}\label{delta3pt}
 \delta^{(4)}\Big(\sum_{i=1}^3 \epsilon_i \omega_i q_i\Big)=\frac{1}{4\omega_3^2} \frac{{\rm sgn}(z_{23} z_{31})}{z_{23} z_{31}} \delta\Big(\omega_1-\frac{\epsilon_3}{\epsilon_1} \frac{z_{23}}{z_{12}} \omega_3\Big) \delta\Big(\omega_2-\frac{\epsilon_3}{\epsilon_2} \frac{z_{31}}{z_{12}} \omega_3\Big) \delta(\bz_{23}) \delta(\bz_{31})\,,
\end{equation}
where I assumed $z_{ij}\neq0$.
The celestial amplitude is
\begin{equation}\label{celestialH3}
 \badat{2}
  \widetilde{\H}_{--+}(\lambda_i;z_i,\bz_i)&=  {\rm sgn}(z_{23}z_{31}) \frac{z_{12}^6}{z_{23}^3z_{31}^3} \delta(\bz_{23}) \delta(\bz_{31}) \left(\frac{\epsilon_2}{\epsilon_1}\frac{z_{23}}{z_{12}}\right)^{i\lambda_1+2}  \left(\frac{\epsilon_3}{\epsilon_1}\frac{z_{31}}{z_{12}}\right)^{i\lambda_2+2}\\
 &\times  \Theta\Big(\frac{\epsilon_3}{\epsilon_1} \frac{z_{23}}{z_{12}}\Big) \Theta\Big( \frac{\epsilon_3}{\epsilon_2} \frac{z_{31}}{z_{12}} \Big) 
  \int_0^\infty d \omega_3 \omega_3^{i\sum_{i=1}^3 \lambda_i}\,.
 \eadat
\end{equation}
Writing it in the form~\cite{Stieberger:2018edy}
\begin{equation}\label{celestialH3temp}
  \widetilde{\H}_{--+}(\lambda_i;z_i,\bz_i)= \frac{{\rm sgn}(z_{23}z_{31})\delta(\bz_{23}) \delta(\bz_{31})}{ z_{12}^{-2+i(\lambda_1+\lambda_2)} z_{23}^{1-i\lambda_1} z_{31}^{1-i\lambda_2}} \Theta\Big(\frac{\epsilon_3}{\epsilon_1} \frac{z_{23}}{z_{12}}\Big) \Theta\Big( \frac{\epsilon_3}{\epsilon_2} \frac{z_{31}}{z_{12}} \Big)\int_0^\infty d \omega_3 \omega_3^{i\sum_{i=1}^3 \lambda_i}\,,
\end{equation}
we see that the celestial amplitude has conformal transformation properties of a three-point correlation function of primary
conformal fields with weights\footnote{For fixed spin $h_i-\bh_i$ we may determine the $\bh_i$ from the $h_i$.}
\begin{equation}\label{hgraviton}
\badat{3}
&h_1=-\frac{1}{2} + \frac{i\lambda_1}{2}\,, & \quad & \bh_1=\frac{3}{2} +\frac{i\lambda_1}{2}\,,\\
&h_2=-\frac{1}{2} +\frac{i\lambda_2}{2}\,, & \quad & \bh_2=\frac{3}{2}+\frac{i\lambda_2}{2}\,,\\
&h_3=1-\frac{i(\lambda_1+\lambda_2)}{2}\,, & \quad & \bh_3=-1-\frac{i(\lambda_1+\lambda_2)}{2}\,.
\eadat
\end{equation}
in agreement with $J_1=-2$, $J_2=-2$, $J_3=+2$ and $\Delta_1=1+i\lambda_1$, $\Delta_2=1+i\lambda_2$. If we analytically continute $\lambda_3$ by shifting it by $i$, namely $\lambda_3=\lambda_3'+i$ with $\lambda_3' \in \mathbb{R}$, then we can interpret the energy integral in~\eqref{celestialH3} as a distribution similar to the case of Yang-Mills amplitudes~\cite{Pasterski:2017ylz,Stieberger:2018edy}:
\begin{equation}
 \int_0^\infty d \omega_3 \omega_3^{i (\lambda_1+\lambda_2+\lambda_3')-1}=
  2\pi \delta\Big(\lambda_1+\lambda_2+\lambda_3'\Big)\,.
\end{equation}
This imposes $\lambda_3=i-(\lambda_1+\lambda_2)$ and therefore yields
\begin{equation}\label{hgraviton3}
\badat{1}
 &h_3=\frac{3}{2}+\frac{i\lambda_3}{2}\,, & \quad  & \bh_3=-\frac{1}{2}+\frac{i\lambda_3}{2}\,,\quad
\eadat
\end{equation}
in agreement with $\Delta_3=1+i\lambda_3$. 
The reason for this analytic continuation is not ad hoc. In fact conformal covariance of the $\delta$-functions in~\eqref{celestialH3temp} imposes the constraint
\begin{equation}
\sum_{i=1}^3 \bh_i=\frac{5}{2}+\sum_{i=1}^3 \frac{i\lambda_i}{2}\,,
\end{equation}
which is precisely satisfied for~\eqref{hgraviton3}. As alluded to in the introduction, an example of a graviton with complex $\lambda$ is the conformal primary $\widetilde{h}^2$ discussed in~\cite{Donnay:2018neh} which is related to the 2D stress tensor for 4D gravity~\cite{Kapec:2016jld} and can be understood as the $\lambda=-i$ limit of a general spin~2 conformal primary with dimension $\Delta=1+i\lambda$. Another example is its shadow transform which is a $\Delta=0$ primary with $\lambda=i$.

\subsection{Celestial four-graviton amplitude}\label{celestialH4point}

The MHV four-graviton amplitude is given by
\begin{equation}\label{H4}
 \H_{--++}(p_i)=H_{--++}\, \delta^{(4)}\Big(\sum_{i=1}^3 \epsilon_i \omega_i q_i\Big)\,,
\end{equation}
where the stripped amplitude is
\begin{equation}
\badat{2}
 H_{--++}&= s_{14}\left[A_4(1^-2^-3^+4^+)\right]^2+ s_{13}\left[A_4(1^-2^-4^+3^+)\right]^2\\\
 &=-s_{12} A_4(1^-2^-3^+4^+)A_4(1^-2^-4^+3^+)\\
 &=4\epsilon_1 \epsilon_2  \frac{\omega_1^3 \omega_2^3}{\omega_3^2\omega_4^2} \frac{z_{12}^7 \bz_{12} }{z_{13} z_{14} z_{23}z_{24}z_{34}^2 }\,.
\eadat
\end{equation}
Here I used the well-known relation $s_{13} A_4(1^-2^-4^+3^+)=s_{14} A_4(1^-2^-3^+4^+)$ and $s_{12}+s_{13}+s_{14}=0$ which follows from energy-momentum conservation. A convenient way for writing the $\delta$-function for the latter is~\cite{Pate:2019mfs}
 \begin{equation}\label{delta4pt}
 \badat{2}
 \delta^{(4)}\Big(\sum_{i=1}^4 \epsilon_i \omega_i q_i\Big)&=\frac{1}{4\omega_3}  \delta\Big(\omega_1-\frac{\epsilon_3}{\epsilon_1} \frac{z_{23}\bz_{34}}{z_{12}\bz_{14}} \omega_3\Big) \delta\Big(\omega_2-\frac{\epsilon_3}{\epsilon_2} \frac{z_{13}\bz_{34}}{z_{12}\bz_{42}} \omega_3\Big) \delta\Big(\omega_4-\frac{\epsilon_3}{\epsilon_4}\frac{z_{23}\bz_{13}}{z_{42}\bz_{14}}\omega_3\Big)\\
 &\times \delta(z_{12}z_{34}\bz_{13}\bz_{24}-z_{13}z_{42}\bz_{12}\bz_{34})\,,
 \eadat
\end{equation}
where I assumed $z_{ij}\neq 0$.
The celestial amplitude is
\begin{equation}\label{celestialH4}
 \badat{3}
 \widetilde{\H}_{--++}&=\epsilon_1 \epsilon_2 \frac{z_{12}^7 \bz_{12}}{z_{13} z_{14} z_{23} z_{24}z_{34}^2} \left(\frac{\epsilon_3}{\epsilon_1} \frac{z_{23}\bz_{34}}{z_{12}\bz_{14}}\right)^{i\lambda_1+3} \left(\frac{\epsilon_3}{\epsilon_2} \frac{z_{13}\bz_{34}}{z_{12}\bz_{42}}\right)^{i\lambda_2+3} \left(\frac{\epsilon_3}{\epsilon_4}\frac{z_{23}\bz_{13}}{z_{42}\bz_{14}}\right)^{i\lambda_4-2} \\
 &\times \Theta\left(\frac{\epsilon_3}{\epsilon_1} \frac{z_{23}\bz_{34}}{z_{12}\bz_{14}}\right) \Theta\left(\frac{\epsilon_3}{\epsilon_2} \frac{z_{13}\bz_{34}}{z_{12}\bz_{42}}\right) \Theta\left(\frac{\epsilon_3}{\epsilon_4}\frac{z_{23}\bz_{13}}{z_{42}\bz_{14}}\right)  \delta(z_{12}z_{34}\bz_{13}\bz_{24}-z_{13}z_{24}\bz_{12}\bz_{34})\\
 &\times \int_0^\infty d\omega_3 \omega_3^{i\sum_{i=1}^4 \lambda_i+1}\,.
 \eadat
\end{equation}
This expression can be brought into a more familiar form
\begin{equation}
 \widetilde{\H}_{--++}=\epsilon_1\epsilon_2\left(\prod_{i<j}^4 z_{ij}^{\frac{h}{3}-h_i-h_j} \bz_{ij}^{\frac{\bh}{3}-\bh_i-\bh_j}\right) z^{\frac{10}{3}} (1-z)^{-\frac{2}{3}} \delta(z-\bz) \int_0^\infty d\omega_3 \omega_3^{i\sum_{i=1}^4 \lambda_i+1}\,,
\end{equation}
where I introduced the conformal cross ratios
\begin{equation}
 z=\frac{z_{12}z_{34}}{z_{13}z_{24}}\,, \quad 1-z=\frac{z_{14}z_{23}}{z_{13}z_{24}}\,,
\end{equation}
and with similar expressions for $\bz$ and $1-\bz$ replacing $z_{ij}\to \bz_{ij}$.
This representation reveals that the celestial four-graviton amplitude has conformal transformation properties of a four-point correlation function of primary
conformal fields with weights~\cite{Stieberger:2018edy}
\begin{equation}\label{h1234graviton}
\badat{3}
&h_1=-\frac{1}{2} + \frac{i\lambda_1}{2}\,, & \quad & \bh_1=\frac{3}{2} +\frac{i\lambda_1}{2}\,,\\
&h_2=-\frac{1}{2} +\frac{i\lambda_2}{2}\,, & \quad & \bh_2=\frac{3}{2}+\frac{i\lambda_2}{2}\,,\\
&h_3=\frac{3}{2}+\frac{i\lambda_3}{2}\,, & \quad & \bh_3=-\frac{1}{2}+\frac{i\lambda_3}{2}\,,\\
&h_4=\frac{3}{2}+\frac{i\lambda_4}{2}\,, & \quad & \bh_4=-\frac{1}{2}+\frac{i\lambda_4}{2}\,,
\eadat
\end{equation}
in agreement with $\Delta_i=1+i\lambda_i$, $J_1=J_2=-2$ and $J_3=J_4=+2$ and where I followed a similar line of reasoning as above and analytically continued $\lambda_3$ by shifting it by $2i$. This again allows us to write the energy integral in~\eqref{celestialH4} as a distribution yielding the relation $\lambda_3=2i-(\lambda_1+\lambda_2+\lambda_4)$.
While this provides an argument for the existence of classical celestial graviton amplitudes, it is not crucial for the conformally soft graviton theorem for $n\geq 4$ gravitons since we can resort to string theory where graviton amplitudes are well-behaved in the UV.

\section{Conformally soft theorem in gravity}\label{csoftgravity}

Weinberg's soft theorem~\cite{Weinberg:1965nx} is the statement that an $n$-particle scattering amplitude factorizes in the limit where the energy of an external graviton is taken to zero:\footnote{In this section I suppress the explicit dependence of the amplitudes on $z_i,\bz_i$.}
\begin{equation}\label{softfactorization}
\lim_{\omega_n\to0} \H_n(\omega_1,\dots,\omega_i,\dots,\omega_n) =S^{(0)} \H_{n-1} (\omega_1,\dots,\omega_i,\dots,\omega_{n-1})+\dots\,,
\end{equation}
where $\dots$ denote subleading corrections\footnote{Universal formulae for the subleading and sub-subleading soft factors $S^{(1)}$ and $S^{(2)}$ also exist~\cite{Cachazo:2014fwa}.}.
The soft factor $S^{(0)}$, expressed in a spinor-helicity basis, is given by~\cite{Cachazo:2014fwa} (taking the $n$-th graviton to have positive helicity)
\begin{equation}\label{S0npt}
 S^{(0)}={-}\sum_{i=1}^{n-1} \frac{[n i]}{\langle n i\rangle} \frac{\langle xi\rangle \langle y i\rangle }{\langle x n\rangle \langle y n\rangle}=\sum_{i=1}^{n-1}\frac{\epsilon_i\omega_i}{\epsilon_n\omega_n}\frac{\bz_{ni}}{z_{ni}} \frac{z_{xi}z_{yi}}{z_{xn}z_{yn}}\,,
\end{equation}
where $x$ and $y$ refer to reference spinors that are judiciously chosen. 

Celestial amplitudes obey the corresponding {\it conformally soft} factorization 
\begin{equation}\label{Csoftnpt}
  \lim_{\lambda_n\to 0} i\lambda_n\widetilde{\H}_{n}(\lambda_1,\dots,\lambda_i,\dots,\lambda_n) = \sum_{i=1}^{n-1} \widetilde{S}^{(0)}_i\widetilde{\H}_{n-1}(\lambda_1,\dots,\lambda_i-i,\dots,\lambda_{n-1})\,,
\end{equation}
relating the celestial $n$-graviton amplitude~\eqref{celestialHn} to a sum of celestial $(n-1)$-graviton amplitudes with shifted conformal weights via the conformally soft factor (again taking the $n$-th graviton to have positive helicity)
\begin{equation}\label{tildeS0npt}
\widetilde{S}^{(0)}_i= \frac{\epsilon_i}{\epsilon_n} \frac{\bz_{ni}}{z_{ni}} \frac{z_{xi}z_{yi}}{z_{xn}z_{yn}}\,.
\end{equation}
The shift in $\lambda_i$ in the $(n-1)$-graviton amplitudes is explained as follows.
The soft graviton theorem~\eqref{softfactorization} can be understood as the Ward identity for the supertranslation current~\cite{Strominger:2013jfa}. As explained in~\cite{Donnay:2018neh} the OPE of the supertranslation current with an operator $\mathcal{O}_\omega$ in the momentum-basis
\begin{equation}\label{OPEomega}
 P_z \mathcal{O}_{\omega} (w) \sim \frac{\omega}{z-w} \mathcal{O}_\omega(w)\,,
\end{equation}
implies in the conformal basis that its OPE with an operator $\mathcal{O}_{(h,\bh)}$ shifts the conformal weights of the latter:
\begin{equation}\label{OPEhbh}
 P_z \mathcal{O}_{(h,\bh)} (w) \sim \frac{1}{z-w} \mathcal{O}_{(h+\frac{1}{2},\bh+\frac{1}{2})}(w)\,.
\end{equation}
In the language of celestial amplitudes the conformally soft limit is thus expected to relate amplitudes with shifted conformal weights: the explicit appearance of $\omega_i$ in the soft factor~\eqref{S0npt} implies that the $i$th external particle has its conformal dimension $\Delta_i=h_i+\bh_i$ shifted by $1$. This corresponds to the shift in $\lambda_i$ by $-i$ in~\eqref{Csoftnpt}.

\subsection{Conformally soft four-graviton amplitude}\label{csoftgravity4point}

I will now verify the conformally soft factorization~\eqref{Csoftnpt} explicitly for the four-graviton amplitude
\begin{equation}\label{csoft4to3factorization}
   \lim_{\lambda_4\to 0} i\lambda_4\widetilde{\H}_{--++}(\lambda_1,\lambda_2,\lambda_3,\lambda_{4}) = \sum_{i=1}^{3}\widetilde{S}^{(0)}_i\widetilde{\H}_{--+}(\lambda_1,\dots,\lambda_i-i,\dots,\lambda_{3}) \,,
\end{equation}
and derive the conformally soft factor
\begin{equation}\label{tildeS04pt}
 \widetilde S^{(0)}_i=\frac{\epsilon_i}{\epsilon_4}\frac{\bz_{4i}}{z_{4i}} \frac{z_{xi}z_{yi}}{z_{x4}z_{y4}}\,.
\end{equation}

Recalling the celestial four-graviton amplitude~\eqref{celestialH4}, the conformally soft limit~\eqref{csoft4to3factorization} is
\begin{equation}\label{csoft4to3temp}
 \badat{3}
  \lim_{\lambda_4\to 0} i\lambda_4\widetilde{\H}_{--++}&=-\epsilon_1 \epsilon_2 \epsilon_3 \epsilon_4 \frac{z_{12}^7}{z_{13} z_{14} z_{23}^2 z_{34}^2} \, z\, \frac{\bz_{14}\bz_{24}}{\bz_{34}} \left(\frac{\epsilon_3}{\epsilon_1} \frac{z_{23}\bz_{34}}{z_{12}\bz_{14}}\right)^{i\lambda_1+3} \left(\frac{\epsilon_3}{\epsilon_2} \frac{z_{13}\bz_{34}}{z_{12}\bz_{42}}\right)^{i\lambda_2+3}\\
 &\quad \times \Theta\left(\frac{\epsilon_3}{\epsilon_1} \frac{z_{23}\bz_{34}}{z_{12}\bz_{14}}\right) \Theta\left(\frac{\epsilon_3}{\epsilon_2} \frac{z_{13}\bz_{34}}{z_{12}\bz_{42}}\right)  \delta(z_{12}z_{34}\bz_{13}\bz_{24}-z_{13}z_{24}\bz_{12}\bz_{34})\\
  &\quad \times  \lim_{\lambda_4\to 0} i\lambda_4 \left(\frac{\epsilon_3}{\epsilon_4}\frac{z_{23}\bz_{13}}{z_{42}\bz_{14}}\right)^{i\lambda_4-1} \Theta\left(\frac{\epsilon_3}{\epsilon_4}\frac{z_{23}\bz_{13}}{z_{42}\bz_{14}}\right) \int_0^\infty d\omega_3 \omega_3^{i\sum_{i=1}^4 \lambda_i+1}\,, 
 \eadat
\end{equation}
where I made use of the $\delta$-function to replace a factor $\frac{\bz_{12}\bz_{34}}{\bz_{13}\bz_{24}}$ by the conformal cross ratio $z$.
The term in the last line becomes
\begin{equation}
 \lim_{\lambda_4 \to 0} i\lambda_4 \Big( \frac{\epsilon_3}{\epsilon_4}\frac{z_{23}\bz_{13}}{z_{42}\bz_{14}}\Big)^{i\lambda_4-1}\Theta\left(\frac{\epsilon_3}{\epsilon_4}\frac{z_{23}\bz_{13}}{z_{42}\bz_{14}}\right)=\delta\Big(\frac{\epsilon_3}{\epsilon_4}\frac{z_{23}\bz_{13}}{z_{42}\bz_{14}}\Big)\,,
\end{equation}
where I used the identity
\begin{equation}\label{deltaID}
 \delta(x)=\lim_{\epsilon \to 0} \frac{|x|^{\epsilon-1}}{\Gamma(\frac{\epsilon}{2})}\,,
\end{equation} 
and $\Theta(0)=\frac{1}{2}$.
Assuming $z_{23}\neq0$ we can write the $\delta$-function as
\begin{equation}\label{d1}
 \delta \Big(\frac{z_{23}\bz_{13}}{z_{24}\bz_{14}}\Big)= {\rm sgn}(z_{23}z_{24} \bz_{14}) \frac{\bz_{14}z_{24}}{z_{23}}\delta(\bz_{13})\,.
\end{equation}
Assuming further $z_{13}\neq0,z_{24}\neq0,\bz_{34}\neq0$ and using~\eqref{d1} we can write
\begin{equation}
  \delta(z_{12}z_{34}\bz_{13}\bz_{24}-z_{13}z_{24}\bz_{12}\bz_{34})=\frac{{\rm sgn}(z_{13}z_{24}\bz_{34})}{z_{13}z_{24}\bz_{34}}\delta(\bz_{12})\,.
\end{equation}
With this the conformally soft limit becomes
\begin{equation}\label{csoft4to3temptemp}
\badat{2}
  \lim_{\lambda_4\to 0} i\lambda_4 \widetilde{\H}_{--++}&=-\epsilon_1 \epsilon_2 \epsilon_3 \epsilon_4  \,\bz_{14}\, z\, \frac{z_{12}^7}{z_{13} z_{14} z_{23}^2 z_{34}^2} \left(\frac{\epsilon_3}{\epsilon_1}\frac{z_{23}}{z_{12}}\right)^{i\lambda_1+3} \left(\frac{\epsilon_3}{\epsilon_2}\frac{z_{31}}{z_{12}}\right)^{i\lambda_2+3} \left(\frac{z_{42}\bz_{14}}{z_{23}\bz_{13}}\right) \\
  &\quad\times \Theta\left(\frac{\epsilon_3}{\epsilon_1}\frac{z_{23}}{z_{12}}\right) \Theta\left(\frac{\epsilon_3}{\epsilon_2}\frac{z_{31}}{z_{12}}\right) \frac{{\rm sgn}(z_{23}z_{31} )}{z_{23}z_{31}} \delta(\bz_{12})\delta(\bz_{13})\int_0^\infty d\omega_3 \omega_3^{i\sum_{i=1}^3 \lambda_i+1}\,.
\eadat
\end{equation}
Comparing with~\eqref{celestialH3} it is clear that the above expression cannot be expressed simply in terms of $\widetilde{\H}_{--+}$. Instead, the OPE~\eqref{OPEhbh} implies that the conformally soft limit relates four-point amplitudes to three-point amplitudes with shifted conformal weights. The way I represented the momentum-conserving $\delta$-function in~\eqref{celestialH4} corresponds to shifting $(h_1,\bh_1)$ in~\eqref{celestialH3} by $(\frac{1}{2},\frac{1}{2})$. This is accommodated by the imaginary shift $\lambda_1 \to \lambda_1 -i$.
The shifted celestial three-point amplitude is
\begin{equation}\label{celestialH3shift}
\badat{2}
 \widetilde{\H}_{--+}(\lambda_1-i,\lambda_2,\lambda_3)&= {\rm sgn}(z_{23} z_{31})\frac{z_{12}^6}{z_{23}^3z_{31}^3} \left(\frac{\epsilon_3}{\epsilon_1}\frac{z_{23}}{z_{12}}\right)^{i\lambda_1+3} \left(\frac{\epsilon_3}{\epsilon_2}\frac{z_{31}}{z_{12}}\right)^{i\lambda_2+2} \\
 &\times \Theta\left(\frac{\epsilon_3}{\epsilon_1}\frac{z_{23}}{z_{12}}\right) \Theta\left(\frac{\epsilon_3}{\epsilon_2}\frac{z_{31}}{z_{12}}\right) \delta(\bz_{13})\delta(\bz_{23}) \int_0^\infty d\omega_3 \omega_3^{i\sum_{i=1}^3 \lambda_i+1}\,.
 \eadat
\end{equation}
From~\eqref{csoft4to3temptemp} and~\eqref{celestialH3shift} follows the conformally soft factorization~\eqref{csoft4to3factorization} with
\begin{equation}\label{tildeS04ptomega1}
 \widetilde{S}^{(0)}=\frac{\epsilon_1}{\epsilon_4}\frac{\bz_{14}}{z_{14}} \frac{z_{12}z_{13}}{z_{24}z_{34}}=s_{14}\, z \,\left(-\frac{1}{2}\frac{z_{31}}{z_{34}z_{41}}\right)^2\,,
\end{equation}
which corresponds to~\eqref{tildeS04pt} with $x=2$ and $y=3$\footnote{Note that any other choice of $x,y$ is equal to~\eqref{tildeS04ptomega1} upon use of the energy-momentum conserving $\delta$-function of the four-point amplitude~\eqref{delta3pt}.}. As an aside, the term inside the brackets in the second equality is the (conformally) soft factor in gauge theory. 

\subsection{Conformally soft graviton theorem in heterotic string theory}\label{csoftgravity4pointHeterotic}

The UV behavior of amplitudes gets softened in string theory. Four-graviton amplitudes in heterotic string theory differ from those in Einstein gravity~\cite{Stieberger:2018edy}
\begin{equation}
 \H_{--++}^{\rm heterotic}=F_H \H_{--++}\,,
\end{equation}
by the heterotic form factor
\begin{equation}
 F_H(s,t,u) =-\frac{\Gamma(-s)\Gamma(-t)\Gamma(-u)}{\Gamma(s)\Gamma(t)\Gamma(u)}\,,
\end{equation}
where $s,t,u$ are the Mandelstam variables
\begin{equation}
 s=\alpha' s_{12}\,, \quad t=\alpha' s_{23}\,, \quad u=\alpha' s_{13}\,.
\end{equation}
For the celestial four-point amplitude~\eqref{celestialH4} this implies the substitution
\begin{equation}
 \int_0^\infty d\omega_3 \omega_3^{i\sum_{i=1}^4 \lambda_i+1} \to \int_0^\infty d\omega_3 \omega_3^{i\sum_{i=1}^4 \lambda_i+1} F_H (\omega_i) \equiv \widetilde F_H(\lambda_i)\,.
\end{equation}
The Mandelstam variables are conveniently expressed as
\begin{equation}
\badat{3}
 s&=4\alpha' \omega_3^2 \frac{z_{23}\bz_{13}}{z_{42}\bz_{14}}z_{34}\bz_{34}\,,\\
 u&=4\alpha' \omega_3^2 \frac{z_{13}\bz_{34}}{z_{12}\bz_{42}}z_{23}\bz_{23}\,,\\
 t&=4\alpha' \omega_3^2 \frac{z_{23}\bz_{34}}{z_{12}\bz_{14}}z_{13}\bz_{13}\,,
 \eadat
\end{equation}
where I made use of energy-momentum conservation of the four-graviton amplitude.
In the conformally soft limit $\lambda_4 \to 0$, using $\bz_{12}=0,\bz_{13}=0$, we get $s,t,u \to 0$ and so the celestial heterotic form factor reduces to
\begin{equation}
 \widetilde F_H\Big|_{\lambda_4\to 0,\bz_{12}\to0,\bz_{13}\to 0}=  \int_0^\infty d\omega_3 \omega_3^{i\sum_{i=1}^4 \lambda_i+1}\,.
\end{equation}
The argument involving the shift $\lambda_i\to \lambda_i-i$ that leads to the energy integral of the three-point amplitude remains unchanged.
Hence, the heterotic four-graviton amplitude obeys the Ward identity
\begin{equation}
 \lim_{\lambda_4\to 0} i\lambda_4 \widetilde{\H}_{--++}^{\rm heterotic}(\lambda_1,\lambda_2,\lambda_3,\lambda_{4}) =\sum_{i=1}^{3}\widetilde{S}^{(0)}_i \widetilde{\H}_{--+}(\lambda_1,\dots,\lambda_i-i,\dots,\lambda_{3}) \,,
\end{equation}
with the conformally soft factor given by~\eqref{tildeS04pt}.

\subsection{Conformally soft $n$-graviton amplitude}\label{csoftgravitynpoint}
To generalize the argument for conformally soft factorization to $n$-point amplitudes recall that the stripped $n$-point MHV graviton amplitude is~\eqref{HnElvangFreedman}
\begin{equation}
 H_n(1^-2^- 3^+ \dots n^+)= \frac{1}{(-2)^{2n-8}}\left(\frac{\omega_1 \omega_2}{\omega_3\dots \omega_n}\right)^2 \sum_{\mathcal{P}(i_3 \dots i_n)} s_{1  i_n} \left( \prod_{m=4}^{n-1} \beta_m\right) \left( \frac{z_{12}^3}{z_{2 i_3} \dots z_{i_n1}}\right)^2\,.
\end{equation}
A convenient way to express $\beta_m$ is~\cite{Elvang:2007sg}
\begin{equation}\label{betassimple}
 \beta_m= 
 \frac{\langle i_m i_{m+1}\rangle}{\langle 2  i_{m+1}\rangle}  (\langle 21\rangle [1 i_m]+\langle 2 i_{m+1}\rangle [i_{m+1} i_m])\,.
\end{equation}
A formal argument for the conformally soft factorization of the $n$-graviton amplitude, similar to the one given in~\cite{Pate:2019mfs} for Yang-Mills, is as follows.
Using the integral representation of the energy-momentum conserving $\delta$-function
\begin{equation}
 \delta^{(4)}\left(\sum_{i=1}^n \epsilon_i \omega_i q_i\right)=\int \frac{d^4y}{(2\pi)^4} e^{-i\sum_i \omega_i(\epsilon_i y \cdot q_i-i\varepsilon)}\,,
\end{equation}
we can write the celestial $n$-graviton amplitude as
\begin{equation}
\badat{3}\label{csoftntonm1temp}
 \widetilde{\H}_n(1^-2^-3^+\dots n^+) &= \left(\prod_{k=1}^n \int_0^\infty d\omega_k \omega_k^{i \lambda_k} \right)  \delta^{(4)}\left(\sum_{k=1}^n \epsilon_k \omega_k q_k\right)\, H_n(1^- 2^- 3^+ \dots n^+)\\
 &=\left(\prod_{k=1}^n \int_0^\infty d\omega_k \omega_k^{i \lambda_k-J_k} \right) \int \frac{d^4y}{(2\pi)^4} e^{-i\sum_k \omega_k(\epsilon_k y \cdot q_k-i\varepsilon)}\\
 &\quad\times  \frac{1}{(-2)^{2n-8}}
 \left[ \sum_{\mathcal{P}(i_3 \dots i_n)} 4 \epsilon_1 \epsilon_{i_n} \omega_1 \omega_{i_n}  \left( \prod_{m=4}^{n-1} \beta_m\right) \frac{z_{1i_n} \bz_{1i_n}  z_{12}^6}{z_{2 i_3}^2 \dots z_{i_n 1}^2}\right]\,.
\eadat
\end{equation}
The crucial thing to notice is that conformally soft poles arise from terms inside $[...]$ that provide a single factor of $\omega_k$ so that the integral over $\omega_k$ becomes\footnote{Here I used
\begin{equation}
 \int_0^\infty d\omega \omega^x e^{-b \omega}=\frac{\Gamma(x+1)}{b^{x+1}}\,.\nn
\end{equation}}
\begin{equation}
 \int_0^\infty d\omega_k \omega_k^{i\lambda_k-J_k+1}  e^{-i \omega_k(\epsilon_k y \cdot q_k-i\varepsilon)} =\frac{\Gamma(i\lambda_k-J_k+2)}{[i \epsilon_k y\cdot q_k+\varepsilon]^{i\lambda_k-J_k+2}}\,.
\end{equation} 
We see explicitly from the $\Gamma$-function that $\widetilde{\H}_n$ has poles at $\lambda_k$ with $3\leq k \leq n$ (but not in $\lambda_1$ and $\lambda_2$) and the conformally soft limit picks out the residue of this pole. The dependence of the $k$-th term on $z_{ij},\bz_{ij}$ disappears in the $\lambda_k\to 0$ limit as can be seen explicitly from the denominator $[i \epsilon_k y\cdot q_k+\varepsilon]^{i\lambda_k-J_k+2}\to1$. From Weinberg's soft theorem it then follows that the remaining terms support the conformally soft factor~\eqref{tildeS0npt} where $n=k$. 

To illustrate these arguments consider again the celestial four-graviton amplitude
\begin{equation}
 \badat{2}
 \widetilde \H_4(1^-2^-3^+4^+)&=\int \frac{d^4y}{(2\pi)^4}  \left(\prod_{k=1}^4 \int_0^\infty d\omega_k \omega_k^{i \lambda_k}  e^{-i \omega_k(\epsilon_k y \cdot q_k-i\varepsilon)}\right)H_4(1^-2^-3^+4^+)\,,
 \eadat
\end{equation}
with the stripped four-graviton amplitude
\begin{equation}
 H_4(1^-2^-3^+4^+)=\sum_{\mathcal{P}(i_3 i_4)}  \frac{\epsilon_1\omega_1}{\epsilon_{i_4}\omega_{i_4}} z_{1i_4}\bz_{1i_4} \left(\frac{z_{i_31}}{z_{i_3i_4}z_{i_41}}\right)^2  H_3(1^-2^-i_3^+)\,.
\end{equation}
Let's first recall the soft theorem.
The above sum over perumtations $\mathcal{P}(i_3,i_4)$ contains two terms both of which contribute to the soft factor. Using energy-momentum conservation we can express the second term in the sum $(i_3i_4)=(43)$ in terms of the first $(i_3i_4)=(34)$ with an additional factor $(z-1)$ thus yielding Weinberg's soft factor~\eqref{S0npt} for $n=4$ for the choice $x=2,y=3$.

Let's now return to the celestial four-graviton amplitude
\begin{equation}\label{csoft4to3temptemptemp}
 \badat{2}
 \widetilde \H_4(1^-2^-3^+4^+)
 &= z_{14}\bz_{14} \left(\frac{z_{31}}{z_{34}z_{41}}\right)^2 z \int \frac{d^4y}{(2\pi)^4} \int_0^\infty d\omega_4 \omega_4^{i\lambda_4-J_4+1} e^{-i \omega_4(\epsilon_4 y \cdot q_4-i\varepsilon)} \\
 &\quad \times \prod_{i=1}^3  \int_0^\infty  d\omega_k \omega_k^{i\lambda_k} \omega_1 e^{-i \omega_k (\epsilon_k y \cdot q_k-i\varepsilon)} H_3(1^-2^-3^+)\,.
 \eadat
\end{equation}
The integral over $\omega_4$ provides the pole at $\lambda_4=0$
\begin{equation}
 \int_0^\infty d\omega_4 \omega_4^{i\lambda_4-J_4+1}  e^{-\omega_4(\epsilon_4 y \cdot q_4-i\varepsilon)} =\frac{\Gamma(i\lambda_4-J_4+2)}{[i \epsilon_4 y\cdot q_4+\varepsilon]^{i\lambda_4-J_4+2}}\stackrel{\lambda_4\to 0}{\longrightarrow} \frac{1}{i\lambda_4}\,,
\end{equation}
which is canceled by the explicit factor $(i\lambda_4)$ in taking the conformally soft limit~\eqref{csoft4to3factorization}. The additional factor of $\omega_1$ in~\eqref{csoft4to3temptemptemp} can be absorbed by shifting $\lambda_1 \to \lambda_1-i$ in the celestial three-graviton amplitude $\widetilde{\H}_3(1^-2^-3^+)$. Hence, the conformally soft theorem~\eqref{csoft4to3factorization}-~\eqref{tildeS04pt} holds.

For higher-point amplitudes it is convenient to use Hodges' formula~\cite{Hodges:2011wm,Hodges:2012ym,Nguyen:2009jk} for the $n$-graviton amplitude which makes Weinberg's soft limits manifest. For example, the stripped five-point amplitude can be expressed as~\cite{Cachazo:2014fwa}
\begin{equation}\label{H5}
 H_5(1^-2^-3^+4^+5^+)=-\sum_{i=1}^4 \frac{[5i]}{\langle 5 i\rangle} \frac{\langle xi\rangle \langle y i \rangle}{\langle x 5\rangle \langle y 5\rangle}H_4(1^-2^-3^+4^+)+\frac{\langle 12\rangle^6}{\langle 23 \rangle^2 \langle 31\rangle^2} \frac{[53][54]\langle 31 \rangle \langle 32 \rangle}{\langle 53\rangle \langle 54\rangle \langle 41\rangle \langle 42 \rangle}\,,
\end{equation}
where the second term corresponds to the subleading soft graviton operator acting on the stripped four-point amplitude $S^{(1)} H_4(1^-2^-3^+4^+)$. With the first term providing the the desired factor $1/\omega_5$, the integral over $\omega_5$, following the same line of reasoning as above, provides the pole at $\lambda_5=0$ of the five-point amplitude in the conformally soft limit. After accounting for the shift in conformal weights of the four-point amplitude, the factor in~\eqref{H5} becomes the conformally soft factor~\eqref{tildeS0npt} for $n=5$. Similar arguments can be applied to the six-graviton amplitude. A pleasant feature of amplitudes with $n\leq 6$ gravitons is that the soft factorization is exact while for higher-point amplitudes there are corrections. While I have not proven the conformally soft factorization~\eqref{csoftgravitynpoint} for general MHV $n$-graviton amplitudes, the general argument given in the beginning of this section combined with the formal arguments described above are highly suggestive.

\section*{Acknowledgements}
I am grateful to Agnese Bissi, Laura Donnay, Sabrina Pasterski, Monica Pate, Ana-Maria Raclariu, Andy Strominger and Ellis Yuan for valuable discussions and for comments on the draft. 
This work was supported by the Black Hole Initiative at Harvard University, which is funded by a grant from the John Templeton Foundation.

\bibliographystyle{style}
\bibliography{references}

\end{document}